# An Embedded Model Estimator for Non-Stationary Random Functions Using Multiple Secondary Variables

Colin Daly[1]

**Abstract** An algorithm for non-stationary spatial modelling using multiple secondary variables is developed herein, which combines geostatistics with quantile random forests to provide a new interpolation and stochastic simulation. This paper introduces the method and shows that its results are consistent and similar in nature to those applying to geostatistical modelling and to quantile random forests. The method allows for embedding of simpler interpolation techniques, such as kriging, to further condition the model. The algorithm works by estimating a conditional distribution for the target variable at each target location. The family of such distributions is called the envelope of the target variable. From this, it is possible to obtain spatial estimates, quantiles and uncertainty. An algorithm is also developed to produce conditional simulations from the envelope. As they sample from the envelope, realizations are therefore locally influenced by relative changes of importance of secondary variables, trends and variability.

## 1 Introduction

A painting, 'Man Proposes, God Disposes' by Edwin Lanseer (1864), hangs in Royal Holloway University, London. An urban myth has it that a student sitting in front of it during exams will fail, unless it is covered up. Since the 1970s, it has indeed been covered during exams. There is no record, however, of how exam failure rates

✉ Colin Daly
cdaly@slb.com

1   Schlumberger, Abingdon, United Kingdom

have responded to this thoughtful gesture on behalf of the university authorities. The story brings to mind a practice that may still be found in some applications of spatial predictive modelling. A model type is chosen. The parameters, such as the covariance function (or variogram), the trends and the data transformations, are calculated carefully with that model in mind. A prior, if used, will be updated by the inference. The fate of that model when confronted by the real world is often less systematically studied at the outset. As the study continues and the real world imparts more information, the model may perform for a while, but generally comes to a grisly end. Ultimately, we fail the test.

To be specific, one concern is that the ensemble behaviour of the modelled random function might not match observable properties of the data set in some types of application. Auxiliary hypotheses are needed for generative models. One of them, stationarity, which in one form or another is needed to allow inference, can occasionally pose problems for the predicted results. The subsurface reality itself is rarely stationary, and without care, its adoption, particularly for simulated realizations of the model, can cause the results to be overly homogeneous.

It would take a very considerable detour to discuss non-stationarity in any detail. Geological reality is certainly not stationary, as a quick walk along an outcrop will show. Sometimes the domain of interest seems to split into regions with radically different behaviour, and subdivision into similar blocks or zones may be a first step towards stationarity. This is a very common approach taken, for example, by software developers and used systematically by their clients in modelling oil reservoirs. Within each region it is sometimes necessary to fit trends or to use intrinsic models (e.g., Chiles and Delfiner 2012). This approach addresses stationarity in the mean, or in the marginal distributions, of the target variable. Another type of stationarity is in the anisotropy of spatial autocorrelation. Work on this aspect dates back to Haas (1990) and Sampson and Guttorp (1992). A good review of the various techniques including space deformation, convolution, basis functions and stochastic partial differential equations, is provided in Fouedjio (2017). This aspect is not considered in the current paper, and the approaches discussed in Fouedjio may be considered complementary.

In many applications a number of secondary variables are available. These are correlative with the target variable but are known at the locations where an estimate of the target is required. These variables may be directly observed, such as geophysical attributes, or may be implicitly observed geometric variables, such as spatial location or distance to the nearest fault. The relationship between variables may be non-stationary. For example, seismic data may vary in quality locally and so their relationship with the target variable is non-stationary. In many generative models, such as Gaussian co-simulation, a simplified relationship, usually linear and stationary, between the covariates and the target variable is assumed, although other possibilities exist. Kleiber and Nychka (2012) allow range, smoothness, anisotropy and variance to vary locally in a multivariate Matérn model. Nonetheless, with the restrictions required for these parametric models, it is difficult to handle real-world situations such as when the target variable is approximately stationary but a secondary variable is not, or vice-versa, or they have different trends but are clearly related in some sense. Note that the classic model distinguishes between spatial location variables, which are called trends, and other covariates. This distinction is not required in the current paper.

Finally, a region which is modelled as stationary may in fact have some subtle, but distinct, sub-regions which affect the model. An example is given in Daly (2021), where an unknown latent variable (facies in that example) can cause the 'true' conditional distribution of porosity to radically change locally. In that example the conditional distribution is sometimes bimodal—when the facies could be one of two with very different porosity distributions—and sometimes unimodal. It is difficult to model such behaviour with the usual stationary parametric models such as the multigaussian. A nested approach, with an explicit modelling of the latent variable, is a tempting solution. However, in geological situations, the spatial law of the implicit model (facies) is very poorly known, and typically very complex. Construction of such a model adds significant complexity which, considering its very high uncertainty, is not always justifiable in terms of effort, clarity, prediction quality or robustness, but is considered by many to be unavoidable, as the univariate conditional distributions that the multigaussian model delivers ultimately produce answers that look unrealistic to a geologist.

Many developments of new algorithms focus on solving a new problem or offering a technical improvement on an existing problem. While this is, of course, a concern in the current paper, the author's experience in delivering software to end users adds a different angle. Let us call a software solution for a project fragile if its correct application requires knowledge, experience or time that the user is unlikely to possess at the moment they need to apply it. The previous paragraphs indicate that using a classic approach for a large three-dimensional spatial modelling project, with many zones, local non-stationarity, multi-model distributions and short deadlines, can very easily become fragile, with the risk that the delivery does not provide a robust solution to the project. One of the motivating factors behind this work was to try to reduce fragility in industrial applications of geostatistical modelling through simplification of workflow and ease of use.

In this article, a simple alternative procedure is proposed which is aimed at reducing these effects. Influenced by the idea of conditional random fields (CRF) and the classic geostatistical wariness of very highly parametric models, the idea is to (partly) estimate the conditional distributions directly based on the secondary data and on prior speculative spatial models. These distributions can be quite non-stationary, reflecting local heterogeneity patterns. They provide estimates of the target variable as well as a locally adapted estimate of uncertainty. This can be done without invoking a full multivariate distribution. Unfortunately, it is not possible to extend the workflow to produce realizations of a random function without such an assumption. In most traditional formulations, this is made up front. Examples are Gaussian random fields (GF), Markov random fields (MRF), hidden Markov models (HMM) and many variants. Since the prior model is generally made with some hypothesis of stationarity, the risk of this hypothesis persisting into the results should be considered.

For the approach considered here, the realizations are made by sampling from the distributions that have been estimated. However, it is only the sampling that needs to be assumed as stationary. Hence, the fully spatial model is only required for the sampling function rather than as a direct model of the earth itself. To summarize:

(1) A machine learning/conditional distribution estimation algorithm is used to obtain a distribution at each spatial location. This family of distributions is called the envelope in this paper.
(2) A stationary ergodic random function is used to sample from the envelope at each location.

The final results are constrained by the results of the estimation phase and may perfectly well turn out to be non-stationary in the usual sense. It turns out to still be possible to make inferences about the type of variogram to use for the stationary sampling RF depending on the assumptions about the type of sampling that is to be used.

The idea of using prior spatial models in the estimation of the envelope is what gives rise to the name Ember, which stands for embedded model estimator. This allows the use of an essentially unlimited number of additional variables to enhance conditioning of the model. Ember works, firstly, by a novel 'embedding' of classic geostatistical techniques into a decision forest-based tool to provide an integrated, consistent estimation of the conditional distribution of the target variable at every location of the field. Secondly, these conditional distributions are used to provide estimates of its mean, quantiles, robust estimates of uncertainty and unbiased estimates of 'sweet spots' such as locations where porosity is expected to be higher than a user-set threshold. It is also used as the basis of stochastic simulations when further conditioning to the hard data is applied. This methodology admits of a number of application strategies. One promising practical strategy, applied in this paper, is to use it as a fast method to provide a good, consistent model of the reservoir with minimal classic geostatistical modelling effort. This is possible because the only essential component of geostatistics needed in the new method is the information provided about the spatial continuity/correlation of the target variable. The solution adopted here is to 'embed' one or more prior models of the correlation into the new algorithm. The algorithm produces a non-linear combination of this information with the additional variables used in property modelling, such as spatial trends and seismic attributes, as well as any number of user-chosen variables that are less frequently used in standard geostatistical methods, such as distance to fault and true vertical depth (TVD). Another major advantage of this method is that the final estimates are now allowed to be non-stationary. In other words, the predictor may 're-weight' the importance of the variables in different parts of the field to adapt to local heterogeneity patterns.

In the next section, the model is developed, although proofs are postponed to the appendix. This is followed by some simple examples, the purpose of which is to show elements of the behaviour of the algorithm. A more detailed realistic synthetic example in three dimensions is to be found in Daly (2021), and for a real case study see Daly et al. (2020).

## 2 Methodology

### 2.1 Conditional Distribution Estimation with Embedded Models

The envelope in Ember can be thought of as an extension of the trends that are used in classic geostatistical models. In the classic model, the trend is a point property (i.e., there is one value at each target location) and is often considered to be an estimate of the mean or of the low-frequency component of the model. It is not exact, in the sense that it does not honour the observed values of the target variable. Typically, it is constructed either as a local polynomial of the spatial coordinates (universal kriging) or using some additional variable(s) (e.g., external drift). In the Ember model, a conditional distribution is estimated at each location. In analogy with the trend, the conditional distribution is built using the spatial coordinates and additional variables. In addition, the envelope estimation step will often use the predictions of simpler models to aid the calculation of the conditional distributions at each location by embedding. In this paper the embedded model is kriging. Models such as kriging can contain information that is not explicitly available in point data through the variogram, which gives information about the lateral continuity of the variable modelled. In the examples it will be seen that including the additional information which kriging brings can help constrain the envelope. Depending on the case, it may be a weak variable, contributing little, or a very strong variable which dominates the solution. Training with embedded models requires a slightly different technique from that using data alone.

In an essay on the methodological underpinnings of spatial modelling, Matheron (1988) discusses the importance of objectivity in model construction. He emphasizes the risk in trying to infer the full multivariate spatial distribution, noting that the over-specified models that we use in practice all lead to relatively simple expressions for the conditional expectation, and that these are far simpler than would be obtained after the fact. It is in the spirit, rather perhaps than the exact letter, of his discussion that this paper attempts to use the availability of multiple secondary data to calculate the envelope and avoid a full estimate of the probability law until it is absolutely necessary for stochastic simulation of the target variable.

In a similar vein, the conditional random fields (CRF) approach avoids construction of the multivariate law (Lafferty et al. 2001). The advantage in direct estimation of each conditional distribution in the envelope compared to a generative Bayesian model is that no effort is expended on establishing relations between the numerous predictor variables. In a full spatial model, these involve stringent hypotheses such as the stationarity of the property of interest (perhaps coupled with some simple model of trend) and the stationarity of the relationship between the target variables and the explanatory variables (e.g., the hypothesis that the relationship between porosity and seismic attributes does not change spatially). One might object that our embedded models are constructed with such hypotheses. This is true, but their influence is mitigated in two ways. Firstly, as shall be seen, the Markov-type hypothesis used removes any direct influence of the construction of the embedded model, instead weighing its influence on the final estimate in an entirely symmetric way with the secondary variables, namely on their ability to predict the target distribution. Secondly, the principal impact of stationarity in the classic model is seen in stochastic realizations which need to invoke the full multivariate distribution and therefore lean heavily on the hypotheses. This can be greatly reduced in the current proposal.

The form of CRF that is used here accommodates and embeds existing spatial models using a Markov-type hypothesis. Let $Z(x)$ be a target variable of interest at the location $x$, and let $\boldsymbol{Y}(x)$ be a vector of secondary or auxiliary variables observed at $x$. Let $\{Z_i, \boldsymbol{Y}_i\}$ be observations of the target and secondary variables observed in the field, i.e. $Z_i$ denotes the value of the target variable $Z(x_i)$ at training location $x_i$. Finally, let $\boldsymbol{Z}_e^*(x) = \boldsymbol{f}(\{Z_i, \boldsymbol{Y}_i\})$, be a vector of pre-existing estimators of $Z(x)$. Then the hypothesis that is required is that the conditional distribution of Z(x) given all available data $\widehat{F}(z|\boldsymbol{Y}(\boldsymbol{x}), \{Z_i, \boldsymbol{Y}_i\})$ satisfies

$$\widehat{F}(z|\boldsymbol{Y}(\boldsymbol{x}), \{Z_i, \boldsymbol{Y}_i\}) = E\big[\mathbb{I}_{Z(x)<z}|\boldsymbol{Y}(x), \{Z_i, \boldsymbol{Y}_i\}\big] = E\big[\mathbb{I}_{Z(x)<z}\big|\boldsymbol{Y}(x), \boldsymbol{Z}_e^*(x)\big]. \quad (1)$$

This hypothesis states that the conditional distribution of $Z(x)$, given all the secondary values observed at $x$ and given all the remote observations of $\{Z_i, \boldsymbol{Y}_i\}$, can be reduced to the far simpler conditional distribution of $Z(x)$ given all the secondary values observed at $x$ and the vector of model predictions at x. The term Markov-type is used in the sense that the sigma algebra $\sigma\big(\boldsymbol{Y}(x), \boldsymbol{Z}_e^*(x)\big)$ on the right of Eq. 1 is coarser than $\sigma(\boldsymbol{Y}(\boldsymbol{x}), \{Z_i, \boldsymbol{Y}_i\})$.

The embedded variables used in this paper are two simple kriging models. Rather than spend time working on the inference of such models, for this paper a long-range and a short-range model are chosen and the contribution of these models is determined together with that of the secondary variables during construction of $\widehat{F}(z|\boldsymbol{Y}(\boldsymbol{x}), \{Z_i, \boldsymbol{Y}_i\})$. This is motivated by the empirical observation that the main contribution of embedded kriging in the new algorithm is to provide information about the lateral continuity of the target variable. This choice allows this simplified version of the Ember estimation process to be fully automated. However, nothing stops the method from being applied without this simplification (and indeed it is not used in some other current applications of the method).

### 2.2 Decision Forests

It is now time to choose an algorithm to estimate the conditional distribution. There are many candidates, but in this paper, the choice made is to base the algorithm on a highly successful non-parametric paradigm, the decision (or random) forest (Breiman 2001, Meinshausan 2006).

A decision forest is an ensemble of decision trees. Decision trees have been used for non-parametric regression for many years, and their theory is reasonably well understood (Győrfi et al. 2002). They work by inducing a partition of space into hyper-rectangles with one or more data points in each hyper-rectangle. As a concrete example, suppose there are observations $(Z_i, \boldsymbol{Y}_i)$, $i = 1, \ldots, n$, with each $\boldsymbol{Y}_i$ being a vector of predictor variables of dimension $p$ and $Z_i$ the $i^{th}$ training observation of a target variable to be estimated. This paper is concerned with spatial data, so with estimation or simulation of $Z(x)$ at a location $x$. The notation $Z_i$ is thus shorthand for $Z(x_i)$, the value of the target variable at the observed location $x_i$. The basic idea behind decision trees is to construct the tree using the training data $(Z_i, Y_i)$ and to predict at a new location x by passing the vector $Y(x)$ down the tree reading off the values of $Z_i$ stored in the terminal node hyper-rectangle, using them to form the required type of estimator.

A decision tree is grown by starting with all data in the root node. Nodes are split recursively using rules involving a choice about which of the $p$ coordinates/variables $Y^p$ in $Y$ to use as well as the value of that variable used for the split, $s_p$, with vectors $(Z_i, Y_i)$ going to the left child if $Y_i^p < s_p$, and to the right child if not. When tree growth terminates (another rule about the size of the terminal node), the geometry of the terminal node is therefore a hyper-rectangle. The associated set of $Z_i$ are stored in that terminal node. In random forests, the choices of variables and splits are randomized in ways explained in the next paragraph, but for now let $\theta$ be the set of parameters chosen and call $T(\theta)$ the tree created with this choice of parameters. Denote by $R$ the hyper-rectangles associated with the leaves of the tree, and call $R(y, \theta)$ the rectangle which represents the terminal node found by dropping $Y(x) = y$ down the tree $T(\theta)$. As stated before, estimations will be created from these terminal nodes, so for example, the estimate of the mean value of $Z(x)$ given $Y(x) = y$, from the single decision tree $T(\theta)$ is

$$\widehat{\mu}(y, \theta) = \sum_{i \in R(y,\theta)} \omega_i(y, \theta) Z_i$$

where the weight is defined by $\omega_i(y, \theta) = \frac{\mathbb{I}_{\{Y_i \in R(y,\theta)\}}}{\sum_j \mathbb{I}_{\{Y_j \in R(y,\theta)\}}}$, i.e., the weight is 1 divided by the number of data in the hyper-rectangle when $Y_i$ is in the terminal node and is 0 when it is not.

Decision forests are just ensembles of decision trees. The random forest is the original type of decision forest introduced by Breiman (2001, 2004), and it has been particularly successful in applications. It combines the ideas of bagging (Breiman, 1996) and random subspaces (Ho 1998). In most versions of decision forests, randomization is performed by selecting the data for each tree by subsampling or bootstrapping and by randomizing the splitting. The parameter $\theta$ is extended to cover the data selection as well as splitting. An adaptive split makes use of the $Z_i$ values to help optimize the split. A popular choice is the CART criterion (Breiman et al. 1984), whereby a node $t$ is split to $t_l = \{i \in t; Y_i^p < s_p\}$ and its complement $t_r$. The variable $p$ is chosen among a subset of the variables chosen randomly for each split such that the winning split variable and associated split value are the ones that minimize the within-node variance of the offspring. A slightly more radical split is employed in this paper. Instead of testing all possible split values for each variable, a split value is chosen at random for each variable candidate and then, as before, the within-node variances for the children are calculated and the minimum chosen. This additional randomness seems to help reduce visual artefacts when using a random forest spatially.

The weight assigned to the $i$th training sample for a forest is $\omega_i(y) = \frac{1}{k} \sum_{j=1}^{k} \omega_i(y, \theta_j)$, where k is the number of trees. The estimator of the conditional distribution needed for Ember is then (Meinshausen 2006)

$$\widehat{F}(Z(x) = z | Y = y) = \sum_{i=1}^{n} \omega_i(y) \mathbb{I}_{\{Z_i < z\}},$$

and the forest estimate of the conditional mean is $\widehat{\mu}(x|y) = \sum_i \omega_i(y) Z_i$.

There is a large body of literature on random forests, so only a few that were directly helpful to the current work are referenced. In addition to the foundational papers by Breiman and Meinshausen, Lin and Jeon (2006) describe how they can be interpreted as potential nearest neighbors (k-PNN), a generalization of nearest neighbour algorithms, and give some good intuitive examples to help understand the role of weak and strong variables, splitting strategies and terminal node sizes. Establishing consistency is relatively easy for random forests with some reasonable conditions (an example is considered later). However, there has not as yet been a full explanation of how Breiman's forest achieves such good results. Nonetheless, some good progress has been made, extending the basic consistency results. The PhD thesis of Scornet (2015) includes a demonstration of the consistency of the Breiman forest in the special case of additive regression models (that chapter was in conjunction with Biau and Vert). Other results concern Donsker type functional central limit theorem (CLT) results for random forests (Athey et al., 2019, and references therein). Mentch and Zhou (2019) investigate the role of the regularization implicit in random forests to help explain their success.

### 2.3 Embedding Models in Decision Forests

If there are one or more models, $M_j(x)$, $j = 1, \ldots, l$, which are themselves estimators of $Z(x)$, it is possible to embed these for use within a random forest. In spatial modelling, the variogram or covariance function can give information about the spatial continuity of the variable of interest. A model such as kriging (Chiles and Delfiner, 2012), which is based on this continuity, is well known as a powerful technique for spatial estimation. In cases where there are a number of other predictor variables (often called secondary variables in the geostatistics literature), the standard method of combining the information is through linear models of coregionalization (Wackernagel, 2003). The linearity of the relationship between variables can be restrictive, and correlations may not be stationary. These considerations, as well as the possibility of a simplified workflow, have prompted the idea of embedding models, which allows the Ember model to capture most of the power of the embedded model while allowing for better modelling of the relationship between variables and of local variability of the target variable. The embedding is lazy in the sense that rather than looking for potentially very complex interactions between variables, it tries to use the recognized power of random forests to deal with this aspect of the modelling. The goal here is to produce an estimate of the conditional distribution $\widehat{F}(Z|Y, M)$.

In the sort of applications considered here, the $Y$ variable will usually include the spatial location, x, meaning that the forest will explicitly use spatial coordinates as training variables. This means that at each location, an estimate of the conditional distribution is $\widehat{F}_x \stackrel{\text{def}}{=} \widehat{F}(Z(x)|Y(x) = y, M(x) = m)$, with $Y$ a vector of size $p$ and $M$ a vector of size $l$. Note that this is not the full conditional distribution at x; rather it will be treated as a marginal distribution at x that will be sampled from later for simulation. To emphasize that this is not a model of a spatial random function, note that the mean of $\widehat{F}_x$, $\widehat{\mu}_x$, is a trend rather than an exact interpolator.

Training a forest with embedded models requires an extra step. In many cases the embedded model is an exact interpolator, so $M(x_i) = Z_i$. This is the case if the embedded model is kriging. If training were to use this exact value, then the forest would overtrain and learn little about how the model behaves away from the well locations, so this is not a useful strategy. A more promising strategy is to use cross-validated estimates. Let $m_{-i}$ be the cross-validated values found by applying the spatial model at data location $x_i$, using all the training data selected for the current tree except that observed at $x_i$. This leads to the training algorithm.

1. Choose the independent and identically distributed (i.i.d.) parameters $\theta_i$ for the ith tree.
2. Select a subset of data, the in-bag samples, for construction of the ith tree (either by subsampling or bootstrapping).
3. For each model and each in-bag sample, using all other in-bag samples, calculate the cross-validated estimate $m_{-i}$.
4. Construct the ith tree using $(Z_i, Y_i, m_{-i})$.

Since the model has now been 'converted into' a datum at each location $x_i$, to keep the notation simple, it is useful to continue to talk about training with $(Z_i, Y_i)$ but to remember that some of the $Y_i$ are actually the predictions of embedded models using cross-validation. The effects of this are as follows:

1. Each tree is constructed on a different data set (as the embedded models depend on the actual in-bag sample).
2. Since the $m_{-i}$ are not independent, the i.i.d. assumption for $(Z_i, Y_i)$ that is usually made for random forests may no longer be valid.

It is worth noting that the same idea can be used to create decision forests where some variables are not known exactly but remain random variables that can be sampled (one value per tree is required). In this view, the current algorithm is actually modelling the embedded variable as a random variable.

### 2.4 Consistency of the Model

Meinshausen (2006) shows that his quantile random forest algorithm is consistent in that the estimate of the conditional distribution converges to the true conditional distribution as the number of samples increases. Such theorems are not always of great practical importance in standard applications of geostatistics. Nonetheless, using random forests for a spatial application with embedded models in these forests, it is worthwhile showing conditions where the convergence is true for the spatial problem despite the fact that the independence requirement no longer holds. Meinshausen's hypotheses are quite demanding compared to the typical practical usage of random forests, so for example leaf nodes contain many values rather than a very few values as is typical in most implementations. Some of the other references made earlier try to weaken these hypotheses and give slightly stronger results (typically at the cost of being non-adaptive, or by requiring a restrictive choice of splitting). However, in this practice-driven paper the simpler result is acceptable.

Remember that $R(\mathbf{y}, \theta)$ is the rectangle which represents the terminal node found by dropping $Y(x) = \mathbf{y}$, down the tree $T(\theta)$. Let the kth edge of this hyper-rectangle be the interval $J(\theta, \mathbf{y}, k)$. This means that the only non-negative weights from the $\theta^{th}$ tree are samples within distance $|J(\theta, \mathbf{y}, k)|$ of the target variable, Z's $k^{th}$ secondary variable $y^k$. The notion of a strong or weak secondary variable is quite intuitive. A strong variable is one which has a significant contribution in the estimation, while a weak variable does not. However, there is no fully agreed on definition of strong or weak. For the sake of this paper, we call a variable strong if $|J(\theta, \mathbf{y}, k)| \to 0$ as $n \to \infty$. If, for a variable, assumptions 1–3 below hold, then the variable is strong (Meinshausen). Note that this depends only on the sample data set and the construction algorithm for the tree. This is stated in Lemma 1 below.

When discussing stochastic simulation, it will be necessary to consider a probability model. Define $U(x) = F(Z(x)|Y(x))$. This is a uniformly distributed random variable. For stochastic simulations in a later paragraph, U(x) will be modelled as a random function.

The assumptions needed for Theorem 1 are as follows.

1. **Y** is uniform on $[0, 1]^p$. In fact, it is only required that the density is bounded above and below by positive constants. Later it is discussed what this means for embedded models.
2. For node sizes,
   (a) The proportion of observations for each node vanishes for large n (i.e., is o(n)).
   (b) The minimum number of observations grows for large n (i.e., 1/min_obs is o(1)).
3. When a node is split, each variable is chosen with a probability that is bounded below by a positive constant. The split itself is done to ensure that each subnode has a proportion of the data of the parent node. This proportion and the minimum probabilities are constant for each node.
4. $F(z|Y = \mathbf{y})$ is Lipschitz continuous.
5. $F(z|Y = \mathbf{y})$ is strictly monotonically increasing for each **y**.

**Theorem 1** (Meinshausen) When the assumptions above are fulfilled, and the observations are independent, the quantile random forest estimate is consistent pointwise, for each **y**.

**Lemma 1** (Meinshausen) When the first three assumptions hold, then $\max_k J(\theta, \mathbf{y}, k) \to 0$ as $n \to \infty$.

Theorem 1 remains valid when using embedded variables provided the hypothesis holds and the samples can be considered i.i.d. The variables are i.i.d. when $U(x) = F(Z(x)|Y(x))$, are i.i.d. While this is not necessarily the case for embedded variables, it is often quite a good approximation and may justify an embedding approach with general models. However, the hypothesis is not required when using kriging as a embedded variable.

A similar result to Theorem 1, namely Theorem 2, holds when using the random forest as a means of producing spatial estimates of the conditional distribution considered in this paper. In fact. the result is, in some ways, stronger when either the embedded variable is kriging and is a strong variable or when the spatial coordinates x,y,z are strong variables in the sense that it converges in $L^2$ as the number of samples increase, essentially because this is an interpolation type problem. For the demonstration, which is left to the appendix, Assumption 1 must be strengthened to ensure that the sampling is uniform, but Assumption 4 may be weakened. Let $\mathcal{D}$ be a closed bounded subset of $\mathbb{R}^n$.

1a) The embedded data variables in Y are uniform on $[0, 1]^p$. In fact, it is only required that the density is bounded above and below by positive constants. Moreover, the samples occur at spatial locations $S_n = \{x_i\}_{i=1}^n$, such that, for any $\epsilon > 0$, $\exists N$ such that $\forall x \in \mathcal{D}, \exists x_l \in S_n$ with $d(x, x_l) < \epsilon$ for all $n > N$. In other words, the samples become uniformly dense.

4a) $Z(x)$, $x \in \mathcal{D}$ is a continuous mean squared second-order random function with standard deviation $\sigma(x) < \infty$.

**Theorem 2** Let $\mathcal{D}$ be a closed bounded subset of $\mathbb{R}^n$. Let $Z : \mathcal{D} \to L^2(\Omega, \mathcal{A}, p)$ be a continuous mean squared second-order random function with standard deviation $\sigma(x) < \infty$, and if, in addition, assumptions 1a, 2, 3 and 5 hold, then

1. If kriging is an embedded model in the random forest which is a strong variable, then the Ember estimate $\widehat{\mu}(x|\mathbf{y}) = \sum_i \omega_i(\mathbf{y}) Z_i$ converges in $L^2$ to Z(x).
2. For a standard random forest with no embedded variables, then if Z(x) also has a mean function $m(x)$, that is continuous in x, if the forest is trained on the coordinate vector x, and if each component of x is a strong variable, then the forest estimate $\widehat{\mu}(x|\mathbf{y}) = \sum_i \omega_i(\mathbf{y}) Z_i$ converges in $L^2$ to Z(x).

With an estimate of the conditional distribution now available at every target location, it is a simple matter to read off estimates of the mean of this distribution, which will be called the Ember estimate in this paper. When kriging is the strongest variable, the Ember estimate is usually close to the kriging estimate but will typically be better than kriging if the secondary variables are the strongest. It must be noted that the Ember estimate, unlike kriging but in analogy with trend modelling, is not exact. It is also possible to quickly read off quantiles, measures of uncertainty such as spread, P90-P10, and interval probabilities of the form $P(a \leq Z(x) \leq b)$.

### 2.5 Stochastic Simulation

Finally, many applications require a means of producing conditional simulations. Since there is an estimate of the local conditional distribution $\widehat{F}(z|Y)$ at all target locations, a simple method to do that is a modification of an old geostatistical algorithm of P field simulation which (1) honours data at the well locations, (2) allows our final result to track any discontinuous shifts in the distribution (e.g., when zone boundaries are crossed), and (3) follows the local heteroscedasticity observed in the conditional distribution as well as the spatially varying relationship between conditioning variables and target. Since each simulation is exact, if it is required for some reason, it is also possible to get a posterior Ember estimate which is exact at the data locations by averaging simulations.

Defining the sampling variable as $U(x) = F(Z(x)|Y(x) = y)$, the required hypothesis is that the sampling variable is a uniform mean square continuous continuous stationary ergodic random function. This means that the class of random functions that can be modelled are of type $Z(x) = \psi_x(U(x))$, for some monotonic function $\psi_x$, which depends on x. To model $Z(x)$ is simply to make a conditional sample from the envelope of distributions $F(Z(x)|Y(x) = y)$ using $U(x)$ such that the result is conditioned at the hard data locations. While there is no restriction on how the uniform field is constructed, for this section assume that $U(x)$ is generated as the transform of a standard multigaussian random function. So $U(x) = G(X(x))$, with G the cumulative Gaussian distribution function and X(x) is a stationary multigaussian RF with correlation function $\rho(h)$. This leads to the relation between Z and X,

$$Z(x) = \varphi_x(X(x)) \equiv \psi_x(G(X(x))).$$

Expand $\varphi_x^i$ in Hermite polynomials $\eta_i(y)$,

$$Z(x) = \varphi_x^0 + \sum_{i=1}^{\infty} \varphi_x^i \eta_i(X(x)).$$

Notice that $\varphi_x^0$ is the mean of $Z(x)$ and so is the mean found in the Ember estimate of the distribution at x. Define the residual at location x by $R(x) = (Z(x) - \varphi_x^0)/\varphi_x^1$. Note that $E[R(x)] = 0 \,\forall x$ and with $\tilde{\varphi}_x^i = \varphi_x^i/\varphi_x^1$, then

$$Cov(R(x), R(y)) = E[R(x)R(y)] = \rho(h) + \sum_{i=2}^{\infty} \tilde{\varphi}_x^i \tilde{\varphi}_y^i \rho^i(h) \text{ with } h = x - y.$$

Progress with inference can be made by considering the location to be a random variable X. Define $C_R(h) = E_X[Cov(R(X), R(X+h))]$, for a random location X to get

$$C_R(h) = \rho(h) + \sum_{i=2}^{\infty} E[\tilde{\varphi}_X^i \tilde{\varphi}_{X+h}^i]\rho^i(h). \tag{2}$$

Since both $C_R(h)$ and $\varphi_X^i$ are available empirically, this allows a possible route for inference of $\rho(h)$. Of course, similarly to universal kriging, since the envelope distributions $F(Z(x)|Y(x) = y)$ are calculated from the data, there is a circularity in the calculation of the covariance of the residuals, but the process seems to work fairly well in practice. A simplification, avoiding the requirement to calculate the expansions, occurs in the (near) Gaussian case where the Hermite polynomial expansion can be truncated at the first order, leading to $\varphi_X^1 = \sigma_{R(X)}$, and

$$C_R(h) \approx \rho(h), \quad (3)$$

so it can be concluded that $\rho(h)$ is, to first order, the covariance of the rescaled residuals of Ember estimates, $(Z(x)-m(x))/\sigma(x)$, leading to a simple approximate inference in the case that each $Z(x)$ is not too far from Gaussian (with an allowable varying mean m(x) and standard deviation $\sigma(x)$).

To obtain a conditional simulation, it is sufficient to choose a value of $u_i = U(x_i)$ at data location $x_i$ such that the observed target variable data $z_i$ is $z_i = F_{x_i}(u_i)$, where $F_{x_i}$ is the envelope distribution at $x_i$. Now, in general, there is more than one such possible $u_i$, particularly when the envelope distribution is estimated using random decision forests. In fact, any $u \in [u_{low}, u_{high}]$ will satisfy this condition where $u_{low}$, $u_{high}$ can be read off the Ember estimate of $F_{x_i}$. Since in this section u is assumed to be created as a transform of a Gaussian RF, then a truncated Gaussian distribution can be used for the sampling (e.g., Freulon 1992). The simulation process becomes the following.

1. Infer a covariance for the underlying Gaussian X(x) using, for example, Eqs. (2) or (3)
2. For each simulation:
   a. Use a truncated Gaussian to sample values of $u_i$ satisfying $z_i = F_{x_i}(u_i)$.
   b. Construct a conditional uniform field U(x) based on X(x) such that U $(x_i) = u_i$.
   c. Sample from $F(Z(x)|Y(x) = y)$ using U(x) for each target location x.

Finally, in practice, a user will want to know how well the simulation algorithm will reproduce the input histogram. Informally, since the envelope of distributions and hence the simulations are made by resampling the training data target variable, it reproduces the histogram provided some values are not sampled far more frequently than others. This depends on whether or not data and target locations are clustered. If the data are not clustered, the size of the partition hyper-rectangles induced by each tree is the same in expectation, and if the target locations are not clustered, then each hyper-rectangle contains approximately the same number of target locations, so the data are sampled approximately equally for each tree, and averaging over many trees brings us closer to the expectation. Therefore, the histogram of target locations will be close to that of the data. If the data are clustered, then more weight in the final histogram is given to isolated data (i.e., the algorithm performs a declustering).

### 2.6 Performance

A referee has suggested a brief discussion on algorithm performance. A full discussion should be the subject of a more algorithmically focused paper. The code was written to be platform-independent in C++ using the Armadillo library for linear algebra and TBB (Threading Building Blocks) for parallel programming. Both the random forest code and the geostatistical algorithms (kriging and turning bands) were written from scratch for the version used here, although Schlumberger's Petrel geostatistical library will be used in a commercial version for reasons of speed in gridded large model applications. The computational load splits into training the model and running the model. The training part is estimation of cross-validated samples and training of the forest. The cross-validation costs $N_{data} \times n_{tree}$ model estimates (i.e., kriging in this case) for each embedded model. The random forest estimation time is standard and generally fast compared to the cost of cross-validation. To run the model on a domain $\mathcal{D}$ requires an estimate of the model at all locations on the domain followed by a conditional stochastic simulation of the uniform random variables. While it is difficult to give exact breakdowns without considerable additional work, the algorithm as presented trains and runs at total cost roughly between 2.5 and 5 times that of a standard Gaussian simulation on the same grid, depending on the relative importance of the size of training versus target data. For example, a current application on a real field model with 300 wells and 12 million cells using 15 secondary data variables runs at about five times the cost of the standard Gaussian simulation algorithm because of the large number of training data (training is roughly 65% of the CPU time). Additional simulations run at approximately the same cost as a classic Gaussian simulation, as training does not need to be repeated. It should be noted that a run of the model produces, as well as a simulation, several estimates of quantiles, an Ember mean estimate, a standard deviation property and several user-defined exceedance probabilities $P[Z(x) > p]$ at virtually no extra cost.

Memory usage depends on the implementation, but here it requires storage of the trees in the forest, an array for each of the desired results (simulations, quantiles, means, etc.) and some workspace. The implementation does not require all results to be held in memory at the same time and, in principle, can scale to arbitrary size models. For example, the algorithm completed in 55 cpu-minutes on a large synthetic model with 1000 wells on a grid of size $10^9$ cells, outputting 14 full grid properties, which is significantly larger than the memory of the Dell Precision laptop, which has 128 GB RAM and Intel Xeon E-2286 M.

## 3 Applications

Three examples are now considered. The first is a classic Gaussian field where kriging and Gaussian simulation are perfectly adapted to solve the problem. It will be seen that Ember performs approximately as well as the classic model in this situation. The second example deals with a complex yet deterministic spatially distributed property. It is not a realization of a random function, and it will be seen that Ember gives a significantly better solution than models which try to interpret it as stationary. A third, less typical example shows a case where the standard approach is not capable of giving a meaningful answer, but Ember can produce something useful.

An important consideration for the new algorithm is reducing fragility. An example of a realistic synthetic model of a distribution of porosity in a subsurface hydrocarbon reservoir with associated geophysical variables is presented in Daly (2021) as it is a bit too involved to discuss here for reasons of space. Estimation studies of reservoir properties in these types of situations generally involve considerable effort and expertise.

That example shows the gains that Ember can bring ease of use and interpretability of results to the workload.

### 3.1 Example 1

Consider a synthetic case of a Gaussian random function (GRF) $Z(x) = \lambda S(x) + \mu R(x)$, where $S(x)$ is a known smooth random function with Gaussian variogram of essential range 100, and $R(x)$ is an unknown residual random function with a spherical variogram of range 70. Consider two scenarios regarding the number of observations, one where 800 samples are known and another where only 50 samples are known. The objective is, knowing the value of $Z(x)$ at the sample locations and $S(x)$ at all locations, to produce an estimate $Z^*(x)$ and realizations $Z_S(x)$. For this simple problem, it is known that cokriging, $Z^K(x)$, coincides with the conditional expectation, so is the optimal estimator. It is also the mean of a correctly sampled realization from the GRF, $Z_S^K(x)$. This will be compared to estimates and simulations from Ember, $Z^E(x)$ and $Z_S^E(x)$, as well as those from a standard ensemble estimate (i.e., without embedded models). The values for $\lambda$ and $\mu$ used to generate the models are 0.7904 and 0.6069, respectively.

For this example, it is assumed that all variograms are known for the cokriging and Gaussian simulation applications, but that assumption is not used for the Ember application. For both Gaussian and Ember, the distribution is evaluated from the available data. For the Gaussian case, a transform is used prior to simulation. There is no need for transforms in an Ember simulation. The cokriging is performed without transformations. The Ember model, being non-parametric, can only estimate distributions from available data and, in the current version at least, can only produce simulations by resampling the available data (it is not too difficult to add some extra resolution, but that is essentially an aesthetic choice).

Figure 1 shows $Z(x)$ and $S(x)$, respectively. The experimental variogram for

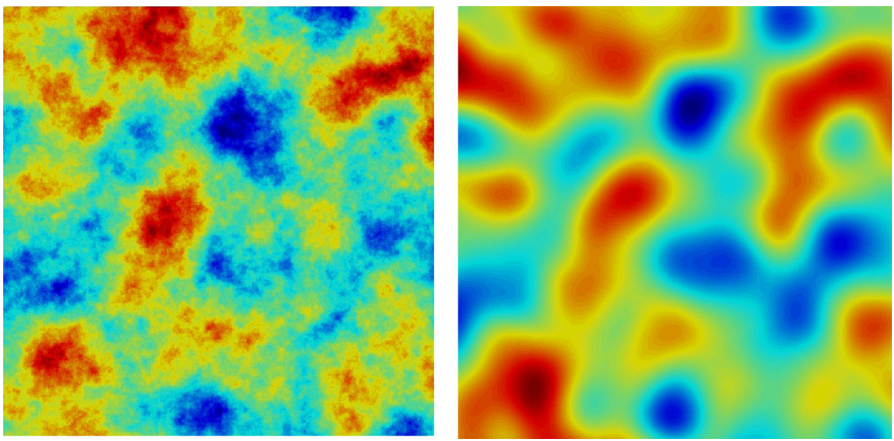

**Fig. 1** L to R, Z(x) and S(x). $Z(x) = \lambda S(x) + \mu R(x)$ for residual R

Math Geoscithe sampling function in the 800-sample case is shown in Fig. 2. As discussed earlier, it is calculated from the relationship for the covariance, $\rho(x - y) = E\left[\left(\frac{Z(x)-M(x)}{\sigma_x}\right)\left(\frac{Z(y)-M(y)}{\sigma_y}\right)\right]$, and has been fitted with an exponential variogram of essential range 34.5. It is shown alongside the experimental variogram of an Ember simulation using that sampling variogram with the model variogram superimposed. The model variogram fits the empirical variogram of the realization quite well.

The results for estimation and simulation for the three models used are shown in Fig. 3, with the estimates on the top row and the simulations on the bottom row. As this is a synthetic model, the true estimation errors can be calculated across all the cells in the image, as well as the errors associated with three simulations from the respective models. These are shown in Table 1. Notice that the Ember model performs very nearly as well as the exact Gaussian method despite using unfitted, standard

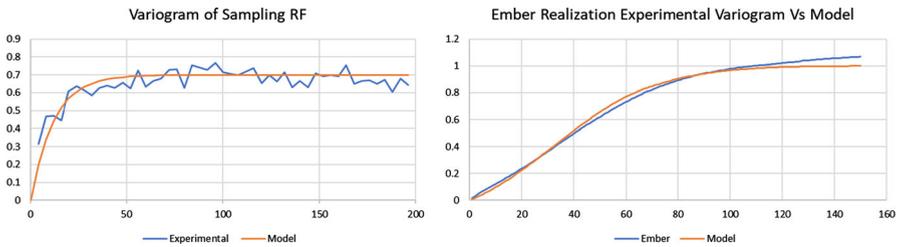

**Fig. 2** Variogram for sampling random function Y in the 800-point case and resulting variogram for the Ember simulation shown in Fig. 3

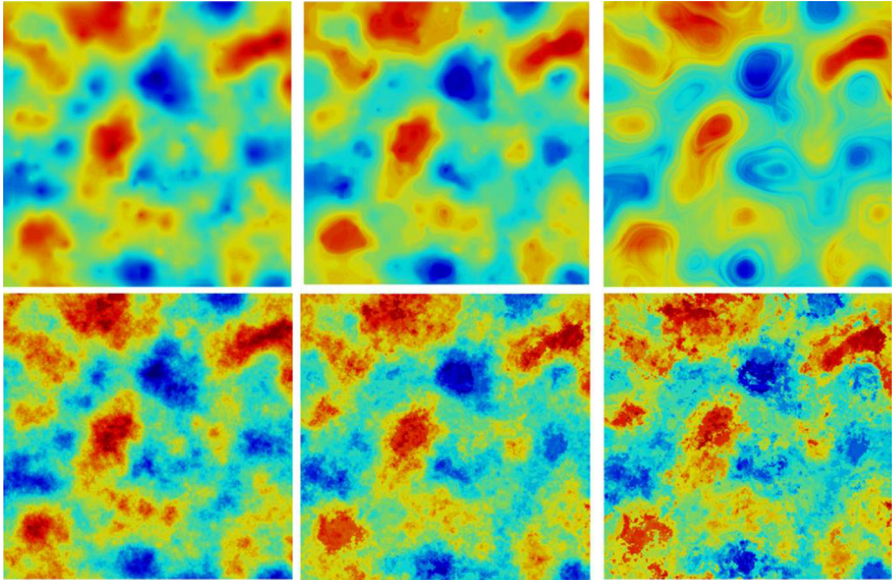

**Fig. 3** L to R Classic Geostats, Ember and ensemble models. Top estimates. Bottom simulations

Springer

**Table 1** Mean squared error (MSE) of estimation and a typical simulation run for the three algorithms

| MSE 800-point case | GRF | Ember | Ensemble |
| --- | --- | --- | --- |
| Estimates | 7.5 | 8.0 | 15.6 |
| Simulations | 15.5 | 16.4 | 29.8 |

embedded models (it is possible to embedded using the correct variogram model, as this is known for this example, but it was interesting to see how it performs in an automated application of the method). The standard ensemble estimate, which does not use embedding, has errors which are about twice as large as the others, showing that embedding kriging models makes a significant contribution in this case. This underlines the argument that significant information can be found and exploited by embedding models as well as using standard secondary data.

Now, turning attention to the case of only 50 samples, the results are shown in Fig. 4 and Table 2, respectively. In this case, only the classic geostatistical results and the Ember results are shown, as the standard ensemble method gives results which are almost identical to the Ember result, because the embedded kriging models are weak variables in this case (the standard ensemble is only marginally better in this case, possibly due to a slight overfitting, but visually almost indistinguishable). This is not surprising, as with so few data, kriging produces highly uncertain estimates, and the secondary trend $S(x)$ has a far higher contribution to the Ember estimate.

The kriging estimate is better than the other two in terms of mean squared error (MSE). This is because (i) The Ember 'estimate' is not an exact interpolator. Rather, it is the mean of the conditional distribution which is a trend. (ii) Kriging is perfectly adapted, as it is equal to the full conditional expectation in the multigaussian case; (iii) the requirement on Ember to estimate the conditional distributions is quite demanding with only 50 samples, though it manages it very well with 800 samples, and (iv) as mentioned earlier, by choice the embedded models used for Ember are the 'standard models' used in the software. In other words, they do not use the correct variogram, so are not adapted exactly to the problem, meaning it is not a like-for-like comparison with the standard geostatistical case. Ember regains ground in simulation, as the Gaussian simulation used a normal score transform so the MSE is slightly more than the theoretical value of twice the kriging MSE. Finally, note that the Ember estimate has less contrast than kriging as it is not an interpolator. The interpolation process in Ember is performed during simulation. An exact Ember estimate can be found by averaging simulations (in petroleum applications, an exact interpolation result alone is not very useful, so this line is not followed in this paper).

To conclude, in this case, where co-simulation is the perfectly adapted optimal method, Ember's results for simulations are effectively as good as those of the classic method. The Ember estimate is not exact, but approaches the optimal algorithm as the number of samples increases.

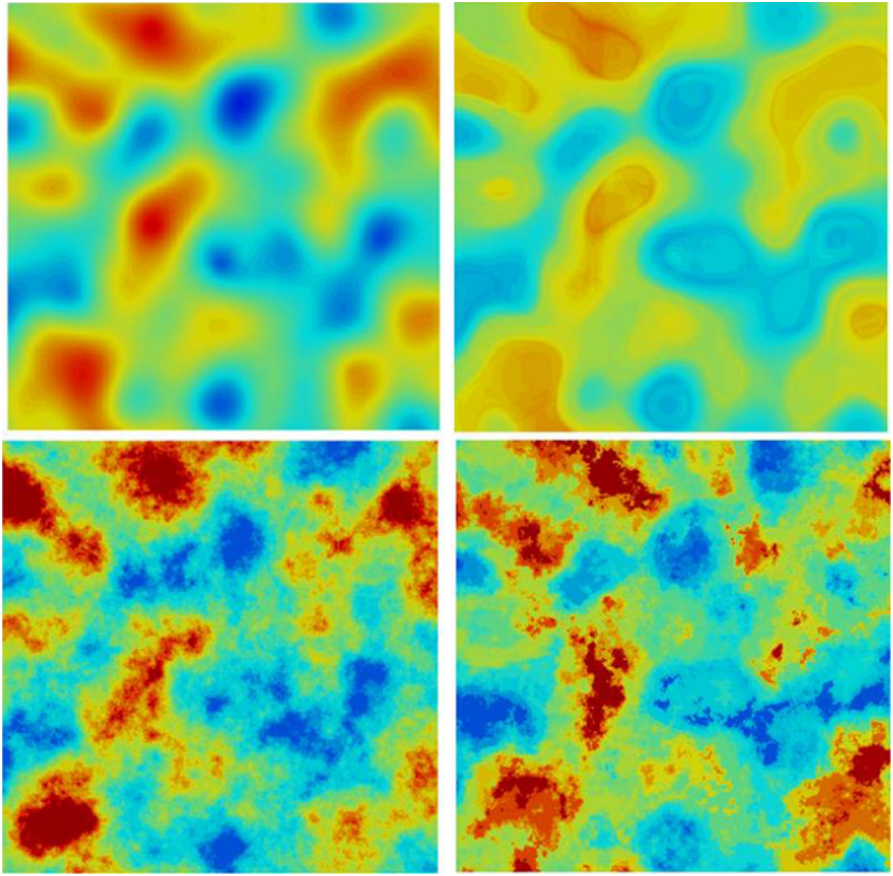

**Fig. 4** 50-point case. Left: Classic Geostats. Right: Ember. Top row are estimates. Bottom row are simulations

**Table 2** First row are the mean squared error (MSE) of estimation for the three algorithms. The second row are MSE of simulations

| MSE 50-point case | GRF | Ember | Ensemble |
| --- | --- | --- | --- |
| Estimates | 27.4 | 36.9 | 35.9 |
| Simulations | 61.1 | 67.6 | 62.4 |

### 3.2 Example 2

This example considers the case where the target variable is generated by a simple deterministic process. After all, most geological processes are 'simple and deterministic'. This case does not resemble any real geological process, but as in many geological processes, interpreting the target variable as a realization of a random field is not at all natural. However, since no knowledge of the process itself is assumed, but only of

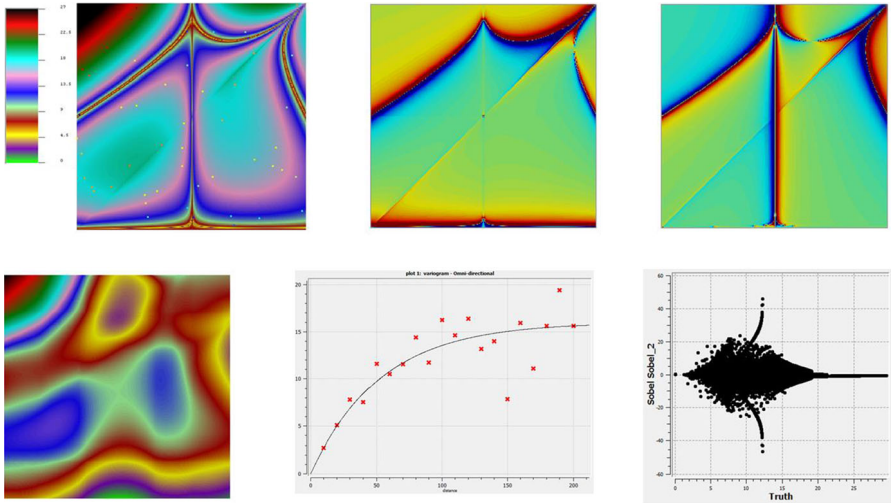

**Fig. 5** Top left, target variable showing 50 sample locations. Other top images are the Sobel filters. Bottom left is the Smooth attribute then the variogram of target variable in 800-point case and on the right, the cross-plot of target with the Sobel filter directly above it the sample data and three secondary variables, it may still be meaningful, and indeed useful, to use stochastic methods to produce estimates. The 'unknown' target variable is shown in Fig. 5 together with the secondary variables. It is a 300 × 300 image which is smooth except at a set of 'cusp' locations where the derivative is not continuous. Consider the two cases, (1) having 800 samples from the image and (2) having just 50 samples from the image (these are just visible as dots on the target image in Fig. 5).

The secondary variables are two Sobel filters of the true target distribution and a heavily smoothed moving average filter of the truth case (filter is a circle of radius 50). An experimental variogram of the target variable in the 800-point case and a cross-plot of the target with one of the Sobel filters are also shown. The cross-plots with the Sobel filters have a zero correlation. The cross-variograms do show some very complex structure, but they are not easily used in linear models of coregionalization.

The objective for this example is to estimate the target variable (hereafter, presumed unknown) using the three secondary variables and the relevant number of training samples.

The results of estimation and simulation for Ember and kriging are shown in Fig. 6. The kriging result for the 50-point case, third from left on top, has a number of bullseyes where the samples fell close to the cusps. This is because a linear model of coregionalization is not easily able to avail of the information contained in the two Sobel filters. It does make good use of the smoothed moving average attribute. Trying to choose models which fit all the data took a bit of time, with a few poor results on the way. On the other hand, the Ember estimate, which is the top left for the 50-point case, is fully automatic, so was obtained with a single 'button click'. In fact, no tinkering was done to any of the Ember estimates in this paper. They all used the same default. All the user changes involve the choice of which attributes to work with. Visually, the

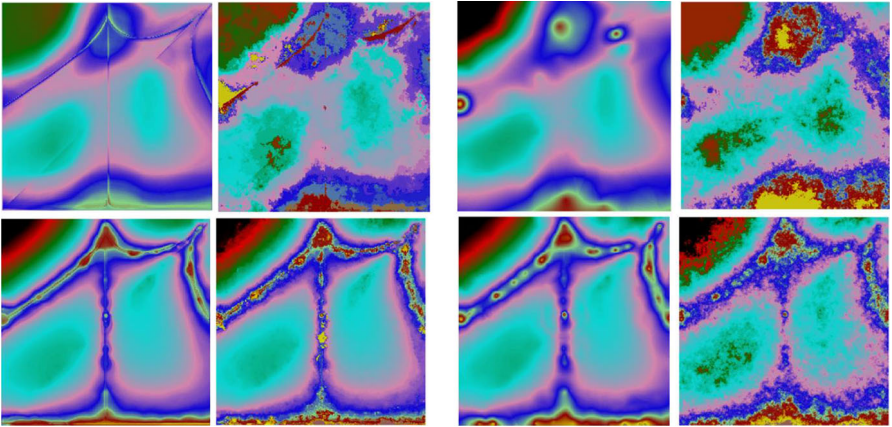

**Fig. 6** Left group of four images are Ember. Right group of four are classic Geostats. In each group, clockwise from top left: estimate 50-point, simulation 50-point, simulation 800-point, estimate 800-point

Ember estimate does a better job at capturing the nature of the target with its cusps, so is more useful for purposes of interpretation. From an MSE perspective, kriging is a BLUE estimator and so is a very good general-purpose estimator. Its results are comparable with Ember, especially with small amounts of data.

The difference becomes more significant in simulation. Even here, the variance of error is not far worse for co-simulation. The issue lies in how these errors are distributed. Remember that what simulation is doing is trying to produce something with the same 'variability' as the target variable. So, loosely speaking, it looks at how much variance has been explained by the estimate of the mean and tops it up until the variability is equal to that of the target. When using GRF with a stationary hypothesis, these errors are spread in a stationary manner (using a Gaussian transform, this is modified, but in a simple way, usually leading to a simple but poorly constrained heteroscedasticity). With Ember, the distribution of error follows the estimated conditional distribution at each location. Looking at the variance of these conditional distributions, Fig. 7 shows that Ember has most of its variability near the cusps, so that the simulations are less variable away from the cusps than in the GRF. The effects are visible in Fig. 6 but are even more clear in Fig. 8, which are images of the relative simulation errors—error divided by the true value—at each location for the 800-point case. These show that the slowly varying areas which have many samples are simulated with a relatively low error by Ember and less well by the other two methods. The cusp regions, which are small and so have fewer sample, are relatively harder to simulate for all methods. This can also be observed by comparing the interquartile ranges (IQR) of the error distributions.

While images for the standard Ensemble method are not shown here, apart from the errors in Fig. 8, Table 3 shows that this method performs as well as Ember in the 50-point case, but significantly worse in the 800-point case. The reason for this can be seen in the measure of variable importance shown in Table 4. In the 50-point case, the dominant variable by far is the smooth attribute, and the embedded models play only a minor role, so that Ember effectively reduces to a standard ensemble, whereas in the 800-point case, the embedded kriging models are most important. So, even though kriging may not be the best adapted model for embedding in this problem because its tendency to smooth is not ideal with the cusps in the target variable, it does almost no harm to the 50-point case and gives a significant gain to the 800-point case.

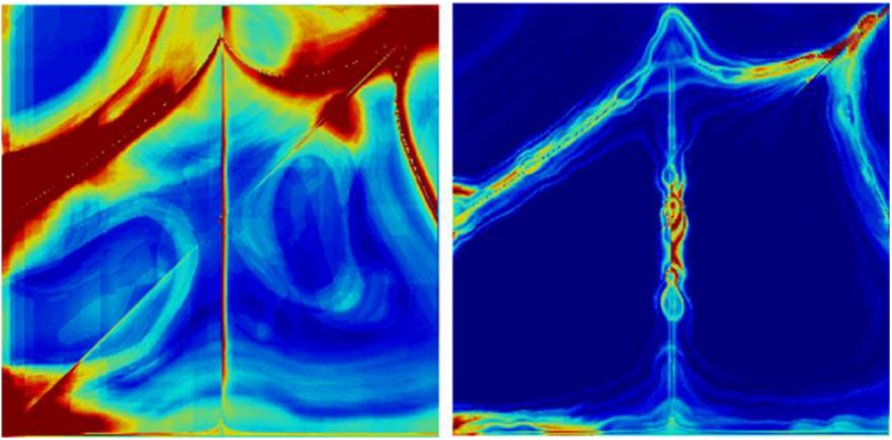

**Fig. 7** Variance of the estimated conditional distribution for Ember. Left is the 50-point case

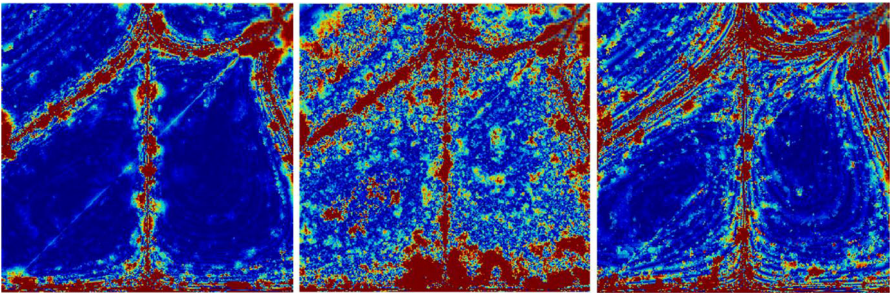

**Fig. 8** Relative errors in simulation, 800 points. L to R, Ember, GRF and simple ensembles

**Table 3** Mean squared error and interquartile range for the estimations and simulations

|          | Cokriging | Ember | Ensemble | GRF Sim | Ember Sim | Ensbl Sim |
|----------|-----------|-------|----------|---------|-----------|-----------|
| 50pt MSE | 3.33      | 3.79  | 3.67     | 7.70    | 6.02      | 5.80      |
| 50pt IQR | 1.33      | 1.23  | 1.18     | 2.99    | 1.91      | 1.84      |
| 800pt MSE| 0.74      | 0.58  | 0.93     | 1.78    | 1.18      | 1.84      |
| 800pt IQR| 0.25      | 0.23  | 0.31     | 1.12    | 0.24      | 0.55      |

Table 4 Variable importance for Ember models. In the case of only 50 data points, the smooth attribute is the strongest variable, but for the 800-point case, the embedded kriging dominates

| Variable importance | | |
|---|---|---|
| | 50-data-point case | 800-data-point case |
| X coordinate | 0.36 | 0.14 |
| Y coordinate | 0.43 | 0.18 |
| Smooth attribute | 15.64 | 2.50 |
| Sobel 1 | 0.61 | 0.43 |
| Sobel 2 | 0.03 | 0.16 |
| Embedded model 1 | 0.85 | 11.49 |
| Embedded model 2 | 0.36 | 7.95 |

Finally, note in passing that a high-entropy Gaussian-based sampling RF is not really well suited to this problem, and it might be interesting to look for lower entropy, or even non-ergodic sampling random functions, to better capture the real nature of the uncertainty of this problem. This is a subject of investigation currently and is not presented here.

### 3.3 Example 3

In this brief example, the target variable is chosen so that standard linear geostatistical models cannot possibly give a good result. The data from example 2 are used but with a new target variable. If in Fig. 9, the variable on the left is called Y(x); then the target variable in the middle is i.i.d random drawn as $Z(x) = N(0, Y(x))$. So, the target variable is a white noise with spatially highly variable standard deviation. Assume that Y(x) is unknown, but run Ember using the same secondary variables as in Fig. 2. While it is possible to generate simulations $Z^s(x)$, as Z is white noise, it will be impossible to make accurate predictions at any given point. However, that does not mean that nothing can be done. It is possible to estimate the local distributions, and hence the local

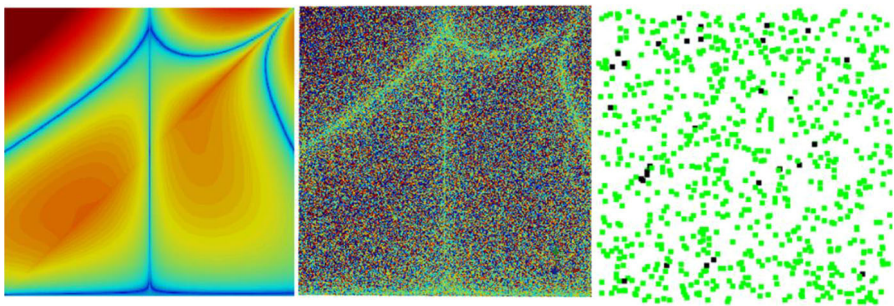

**Fig. 9** The figure in the middle, Z(x), is a pure white noise generated with mean 0 and standard deviation equal to the figure on the left, Y(x). The 'texture' is simply an artefact of the locally changing variance. On the right, the black squares are the samples with value greater than 3

variability. Moreover, the probability that Z(x) exceeds a user-defined critical threshold can be estimated. Choose the threshold to be 3; for Z(x) below, $P(Z(X) > 3) \approx 0.03$, so is in the tail of the distribution. While this example has been generated without a real example in mind, this type of procedure may be useful for problems where individual events are too local and noisy to be predicted individually, but where their density is of importance, such as certain problems involving microseismic, fracture density or in epidemiology.

As before, there are 800 samples to train with. There is no spatial correlation of the variable Z. The set of points with Z(x) > 3 is a point process which *does* cluster, as their location is partly determined by $Y(x)$. The sample of that process for the 800-point case is shown on the right in Fig. 9. Estimation of the density of that process is difficult with only 800 points, as the number of observations of such extreme values is very low—shown in black. In Fig. 10, the upper row shows the estimate of the standard deviation Z(x). It is noisy with only 800 samples but begins to converge with a large number of samples (88,000). In the lower row, the estimate of Prob(Z(x) > 3) with the 800-point case is on the left and the 88,000-point case in the middle. This can be qualitatively, but not rigorously, compared to a smoothed version of the true indicator (using a 5*5 moving average), which is on the right.

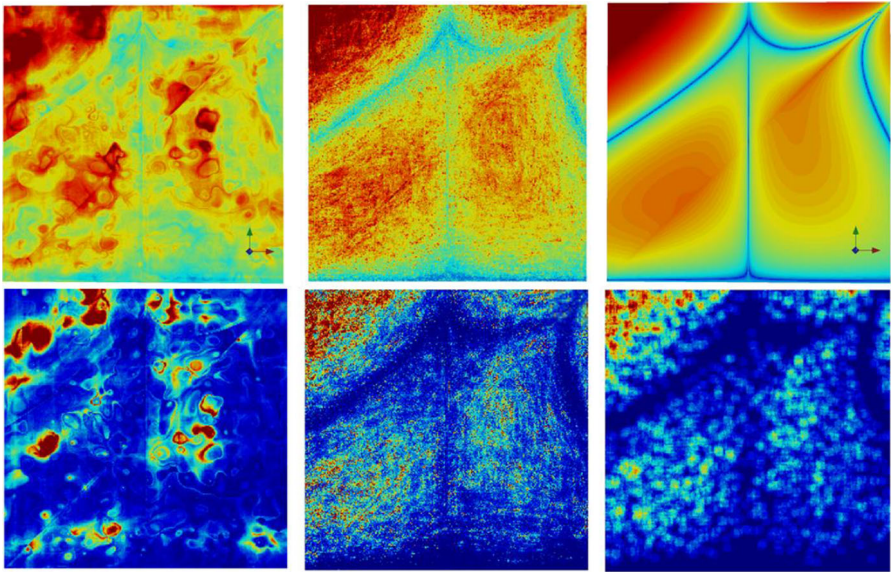

**Fig. 10** Upper Left estimate of the standard deviation of Z(x). Left from 800 samples, centre from 88,000 samples. True value on the right. Lower left is estimate of prob(Z(x) > 3) from 800 samples, centre from 88,000 samples, and on the right a low-pass filter over the true case

## 4 Conclusions

This paper presents an algorithm for spatial interpolation and simulation motivated by the requirements to make better use of all available data and to simplify the modelling process for users. It combines a density estimation algorithm with a spatial estimator to produce an envelope of conditional distributions at each location of the domain $\mathcal{D}$ where results are required. The algorithm used is a decision forest modified so that it trains on the predictive ability of the spatial estimator as well as standard data variables (embedding). Moments or quantiles from the envelope can be used to make deterministic predictions on $\mathcal{D}$. Realizations with realistic geological texture can be performed by sampling from the envelope with an appropriate stationary random function (RF) allowing for additional hard conditioning at the data sample locations if required. The RF plays a somewhat different role here, as it is not used as an explicit model for the geological phenomena of interest, but rather as a way of exploring scenarios constrained by direct estimate of the envelope of distributions. Nonetheless, it is still possible to perform an inference of the covariance function of the sampling RF to serve as a starting point for scenario construction.

## Appendix A

The results below refer to the numbering of the assumptions stated in the main body of the text, prior to the statement of theorems 1 and 2.

**Lemma 2** With a random function satisfying 4a and a sampling following 1a, then for any $\epsilon > 0$, $\exists N$ such that for any $x \in \mathcal{D}$ and for any $n > N$ there is a sample $x' \in S_n$ satisfying $Var\left(Z(x) - Z(x')\right) < \epsilon$.

**Proof** Since $Z(x)$, $x \in \mathcal{D}$ is a continuous mean squared second-order random function, then $C(x, y) : \mathcal{D} \times \mathcal{D} \to \mathbb{R}$ is continuous. $\mathcal{D}$ is compact, so the Heine-Cantor theorem tells us that $C(x, y)$ is uniformly continuous for the product topology on $\mathcal{D} \times \mathcal{D}$. Therefore, for any $\epsilon > 0$, $\exists \delta > 0$, with $\delta$ independent of x, such that for any y, then $|x - y| < \delta \Rightarrow |C(x, y) - C(x, x)| < \epsilon/2$.

Using the above result with the fact that sampling becomes uniformly dense, then for any x, there is an N, independent of x, and a subsequence $\{x_{n_k}\}$, so that for any $n > N$, $|x - x_{n_k}| < \delta$, and so $|C(x, x_{n_k}) - C(x, x)| < \epsilon/2$.

Finally, $Var(Z(x) - Z(x_{n_k})) \leq |C(x, x) - C(x, x_{n_k})| + |C(x_{n_k} x_{n_k}) - C(x, x_{n_k})|$ by the triangular inequality, and since the bound is independent of x, both terms on the right-hand side are less than $\epsilon/2$ when $n > N$, and the result follows.

Assumptions 1 and 3 are used in Meinshausen's proof that the edges of terminal node hyper-rectangles, $J(\theta, y, j)$, become small as the number of samples increases for any variable j used in the standard random forest. The variable j is therefore a strong variable. When using kriging as an embedded variable, this remains true, provided assumption 1 applies for the embedded model. This requirement seems unfamiliar. It states that the set of values that kriging gives when calculated over the domain, $\{z : \exists x \in \mathcal{D} \text{ with } z = z^K(x)\}$, is densely sampled by the set of cross-validated estimates at the data locations. In fact, it states something much stronger, that they sample densely for any configuration of the other secondary variables. This is unlikely to happen in practice with a limited amount of sample data, but then, neither are the requirements for convergence of kriging or a standard random forest for that matter.

**Theorem 2** Let $\mathcal{D}$ be a closed bounded subset of $\mathbb{R}^n$. Let $Z : \mathcal{D} \to L^2(\Omega, \mathcal{A}, p)$ be a continuous mean squared second-order random function with standard deviation $\sigma(x) < \infty$, and if, in addition, assumptions 1a, 2, 3 and 5 hold, then

(1) If kriging is an embedded model and it is a strong variable for the random forest, then the Ember estimate, $\widehat{\mu}(x|y) = \sum_i \omega_i(y) Z_i$, converges in $L^2$ to $Z(x)$.
(2) For a standard random forest with no embedded variables, then if $Z(x)$ also has a mean function, $m(x)$, that is continuous in x, if the forest is trained on the coordinate vector x, and if each component of x is a strong variable, then the forest estimate, $\widehat{\mu}(x|y) = \sum_i \omega_i(y) Z_i$, converges in $L^2$ to $Z(x)$.

**Proof** Start by demonstrating part 1. First note that kriging is consistent. For any $\epsilon > 0$, and for any $x \in \mathcal{D}$, Lemma 2 shows that there is an $N_1$ such that for $n > N_1$, there is a $x' \in S_n$ satisfying $Var\left(Z(x) - Z(x')\right) < \epsilon/9$. Then, for the kriging estimate of $Z(x)$, with $\lambda_i$ being kriging weights,

$$Var\left(Z(x) - \sum_{i=1}^n \lambda_i Z_i\right) \leq Var\left(Z(x) - Z(x')\right) < \frac{\epsilon}{9},$$

using that $Z(x)$ is mean squared continuous. Notice that the same result applies, for the same $\epsilon$, to the kriging cross-validation at any sample (to account for the removal of the validation point, it may be necessary to increase the value of $N_1$ to ensure this holds at all samples).

To show consistency for the Ember estimate, it must be shown that

$$Z(x) - \widehat{\mu}(x|y) \to 0 \text{ in } L^2, \tag{4}$$

The Ember estimate error will be shown to split into three parts, a kriging error at the target location, an error term relating the kriged values at the target location to the cross-validated values at the data locations, and a term containing cross-validation errors at the data locations. The first and third are dealt with by the property of kriging mentioned above. The second term will tend to zero due to the rules of construction of the forest and the fact that kriging is a strong variable. The second term is now considered, before putting it in context later.

Since the forest weights are just the mean of the tree weights $\omega_i(\mathbf{y}) = \frac{1}{k}\sum_{j=1}^{k}\omega_i(\mathbf{y},\theta_j)$, and since the number of trees is fixed and finite, then it is sufficient to show that the result above holds for individual trees. Drop reference to the parameter $\theta_i$, but understand that reference is to an individual tree in what follows.

The tree is a way of constructing the estimator random variable $\hat{\mu}(x|\mathbf{y})$. Each instance of the set of r.v $\{Z_i\}$ will lead to a different tree, but the construction will ensure that the estimator satisfies the required property. First show a bound for any instance of the set of samples of the random variables $\{Z_i = z_i\}$. As before, using a hyphen to indicate excluded values, the cross-validated kriging at sample $i$ will be written as $z_{-i}^K$. Since $\omega_i(\mathbf{y})$ is only non-zero for estimation of $Z(x)$ when sample $i$ is in $R(\mathbf{y},\theta)$, then that sample must have a kriged value within the length of the edge of $R(\mathbf{y},\theta)$ in the kriging variable direction which was denoted earlier by $J(\theta,\mathbf{y},k)$ where k is the index of the embedded kriging variable in Y. So, since kriging is a strong variable and the corresponding edge drops to 0, then if $z_i$ has non-zero weight and if $\epsilon > 0$, then $\exists N_2$ such that

$$\left|z^K(x) - z_{-i}^K\right| < |J(\theta,\mathbf{y},k)| < \frac{\sqrt{\epsilon}}{3} \forall n > N_2.$$

Since $\omega_i(\mathbf{y}) > 0$, $\sum \omega_i(\mathbf{y}) = 1$, then using the triangular inequality

$$\left|z^K(x) - \sum_{i=1}^{n}\omega_i(\mathbf{y})z_{-i}^K\right| < \frac{\sqrt{\epsilon}}{3} \forall n > N_2.$$

But $z^K(x) - \sum_{i=1}^{n}\omega_i(\mathbf{y})z_{-i}^K$ are arbitrary instances of the random variable $Z^K(x) - \sum_{i=1}^{n}\omega_i(\mathbf{y})Z_{-i}^K$. Hence the tree construction ensures that

$$\left|Z^K(x) - \sum_{i=1}^{n}\omega_i(\mathbf{y})Z_{-i}^K\right| < \frac{\sqrt{\epsilon}}{3} \forall n > N_2 \text{ for the random variables } \{Z_i\}. \quad (5)$$

Since this is now a bounded random variable on the interval $\left[-\frac{\sqrt{\epsilon}}{3}, \frac{\sqrt{\epsilon}}{3}\right]$, Popoviciu's inequality on variances implies that $Var\left(Z^K(x) - \sum_{i=1}^{n}\omega_i(\mathbf{y})Z_{-i}^K\right) < \frac{\epsilon}{9}$, for $n > N_2$. To show A.1, it is sufficient to show that $E[Z(x) - \hat{\mu}(x|\mathbf{y})] \to 0$, and $Var[Z(x) - \hat{\mu}(x|\mathbf{y})] \to 0$. Noting that

$$Z(x) - \hat{\mu}(x|\mathbf{y}) = \left(Z(x) - Z^K(x)\right) + \left(Z^K(x) - \sum_{i=1}^{n}\omega_i(\mathbf{y})Z_{-i}^K\right) + \left(\sum_{i=1}^{n}\omega_i(\mathbf{y})Z_{-i}^K - \hat{\mu}(x|\mathbf{y})\right)$$

$$= \left(Z(x) - Z^K(x)\right) + \left(Z^K(x) - \sum_{i=1}^{n}\omega_i(\mathbf{y})Z_{-i}^K\right) + \left(\sum_{i=1}^{n}\omega_i(\mathbf{y})\left(Z_{-i}^K - Z_i\right)\right) \quad (6)$$

then the first and third terms on the right are kriging errors and sums of kriging errors, respectively, so have zero expectation. For the second term, while the expectation $E\left(Z^K(x) - \sum_{i=1}^{n} \omega_i(y)Z_{-i}^K\right)$ is not exactly equal to zero in the general case, (5) shows that it converges in mean of order 1, i.e. $E\left|Z^K(x) - \sum_{i=1}^{n} \omega_i(y)Z_{-i}^K\right| \to 0$. Hence, we can conclude that the expectation of the left-hand side $E[Z(x) - \widehat{\mu}(x|y)] \to 0$.

Also, it has already been shown that the second term has variance less than $\epsilon/9$, and that by uniformity, the first term and all of the individual kriging errors in the third term are all bound by the same value of $\epsilon/9$ for $n > N_1$. To bound the third term fully, expand and use Cauchy-Schwartz

$$Var\left[\left(\sum_{i=1}^{n} \omega_i(y)\left(Z_{-i}^K - Z_i\right)\right)\right] \leq \left(\sum_{i=1}^{n} \omega_i(y)\sqrt{Var\left(Z_{-i}^K - Z_i\right)}\right)^2 < \frac{\epsilon}{9} \forall n > N_1.$$

So, choosing $N = \max(N_1, N_2)$, then the variance of all three terms on the right of (6) is less than the $\epsilon/9$ for $n > N$, and using the well-known simple inequality for variances of square integrable random variables, $Var(\sum_{i=1}^{k} X_i) \leq k \sum_{i=1}^{k} Var(X_i)$, we conclude that

$$Var(Z(x) - \widehat{\mu}(x|y)) < \epsilon \ \forall n > N$$

Since both variance and expectation converge, (A.1) follows.

Part 2 follows along similar lines. As before, it is required to show that (4) converges in mean and that the variance tends to zero. Since the coordinates are strong variables and since $|I(\theta, y, k)|$ becomes arbitrarily small as n increases, it means that all the samples with non-zero weights are close in Euclidean metric to x, so the uniform continuity of the covariance allows us to ensure that, given an $\epsilon$, then $Var[(Z(x) - Z_i)] < \epsilon$, for sufficiently large N. Writing the variance of error in (4) as $Var\left[\sum_{i=1}^{n} \omega_i(y)(Z(x) - Z_i)\right]$, the result follows as before. For the convergence of the mean in (4), the argument previously used the decomposition in (6), but this cannot apply, since this time there is no embedded kriging to ensure that the kriged values are the same. Depending only on the proximity of all the samples to x, the continuity of the mean function m(x) must be used to strengthen the result.

*Note 1* Part 1 shows that a random forest with an embedded kriging will converge to the realization of the random function for uniform dense sampling provided that the embedded kriging is strong. Also, the limiting result does not depend on using the correct variogram in the embedding, although this will help to speed convergence.

*Note 2* Part 2 of the theorem shows that a random forest without embedded models can still converge to the realization of the random function for uniform dense sampling provided that all the coordinate variables are strong.

*Note 3* In practice, it often occurs that both the coordinates and the embedded kriging are strong variables. This accelerates convergence compared to either being strong alone.

*Note 4* The reach of the above theorem extends to other embedded models. As the limit does not depend on the correct choice of covariance, the result can be seen to hold for models that are dual to kriging in an appropriate reproducing kernel Hilbert space, such as interpolating splines or RBF. Also, the proof works with no significant change for any projection type estimator in $L^2(\Omega, \mathcal{A}, p)$. So, with minor adaptation, it works for non-linear estimators such as disjunctive kriging and conditional expectation.

*Note 5* It may be that assumption 1a can be strengthened to ensure that the embedded variable is strong (i.e. it should imply that assumption 1 holds for the extended space of data and embedded variables as then Meinshausen Lemma 2 applies). To get a feel for why, remember that the kriged value needs to be approximated by a cross-validated sample. Consider a kriged value at an arbitrary location, $Z^K(x)$. Since the sampling is dense, there is a sample at some $x_i$ close to $x$ such that for any $\epsilon$, $Var(Z(x) - Z_i) < \epsilon/9$. Since $Z_i$ is used for kriging, $Var(Z(x) - Z^K(x)) < Var(Z(x) - Z_i) < \epsilon/9$, and similarly $Var(Z^K_{-i} - Z_i) < \epsilon/9$. Then decomposing $(Z^K_{-i} - Z^K(x)) = (Z^K_{-i} - Z_i) + (Z_i - Z(x)) + (Z(x) - Z^K(x))$, and applying $Var(\sum_{i=1}^k X_i) \le k \sum_{i=1}^k Var(X_i)$, shows that $Var(Z^K_{-i} - Z^K(x)) < \epsilon$, giving the approximation required. With appropriate regularity conditions on the data variables, this argument can probably be extended to ensure assumption 1 holds on the extended space.

## Declustering?

The convergence theorems are perhaps comforting and give us a glimpse into why the methods work. However, in earth science practice, the amount of data available for training is often extremely limited, particularly in terms of its spatial coverage. Not only can data be quite scarce, but often it is preferentially sampled. In practice, the hyper-rectangles from a decision tree do not actually drop to almost zero in size. Moreover, they tend to be significantly larger in regions of sample space which are sparsely populated. In other words, clustered samples tend to have less influence than sparse samples if the true distribution of the sample is biased. By a heuristic 'wisdom of crowds' type argument, this declustering effect is likely to be enhanced by the fact that a forest makes use of many trees, each with a different partition. This is just a speculation for now but seems worthy of further investigation.

## Appendix B

## Embedding Models in Forests

In this appendix, the idea of embedding models in a forest is looked at more closely. Embedding is done through a cross-validation approach. The objective is to make use of information about the spatial continuity of the target variable. While this does not extract all the information from a theoretical model—an example is given with short-range variability where it is clearly not—it must be remembered that the objective of the forest is only to provide the distribution envelope based on empirical knowledge. Only the part of the information from the embedded model that can be corroborated by the data is used. The short-range variability that cannot be corroborated must then be handled in the simulation/estimation phase. Start with some well-known results for estimators deriving from a projection (in particular, kriging), but with a geometrical proof more suited to current purposes.

Let $E$ be a set and $Z : E \to L^2(\Omega, \mathcal{A}, p)$ be a map from $E$ to a space of square integrable random variables. Typically, $E$ will be a subset of $\mathbb{R}^n$. For $x \in E$, $Z(x)$ will be a random variable. Only the simplest case is considered here, where the covariance is strictly positive definite though not necessarily stationary. Without loss of generality, take $E[Z(x)] = 0 \,\forall x$. Its covariance is $C_{xy} = Cov(Z(x), Z(y)) = <Z(x), Z(y)> = E[Z(x)Z(y)]$. Given a set of data $\{Z_i\}_{i=1}^n$ with $Z_i = Z(x_i)$, let $T_n$ be a complete linear space generated by functions of $\{Z_i\}_{i=1}^n$. The case of greatest interest here is when it is the space of linear combinations $T_n = \text{span}\{\sum_{i=1}^n \lambda_i Z_i\}$. Define $T_n^{-i}$ to be the space generated by all the data variables except $Z_i$, $\mathcal{P}_n^{-i}$ the orthogonal projection onto $T_n^{-i}$ and define $I_i = Z_i - \mathcal{P}_n^{-i} Z_i$, to be the $i$th innovation (i.e., the cross-validation residual). Then we have the following.

**Theorem 3** When $T_n$ is the space of linear combinations, then the orthogonal projection onto $T_n$ is just kriging. Since $C_{ij}$ is strictly positive definite, then its inverse exists, $P^{ij}$ and is known as the precision matrix. Then

a. $<I_i, I_j> = \frac{P^{ij}}{P^{ii} P^{jj}}$

b. The cross-validated result at the ith sample, $Z_{-i}^K$ is given by $Z_{-i}^K = -\sum_{j \neq i} \frac{P^{ij}}{P^{ii}} Z_j$

c. The $\{I_i\}$ span $T_n$ and for any $x \in E$, $Z^K(x) = \sum_i C_{ix} \frac{I_i}{\|I_i\|^2}$

**Proof** Since $C_{xy}$ is non-singular, $\{Z_i\}_{i=1}^n$ form a basis for $T_n$. Moreover, there exists a dual basis $\{Z^i\}_{i=1}^n$ with $<Z^i, Z_j> = \delta_j^i$. Let the transformation matrix between the basis be defined by $Z^i = \sum_{i=1}^n P^{ij} Z_j$. $P^{ij}$ is the precision matrix, and the above results hold. To see this, first note,

$$\delta_j^i = <Z^i, Z_j> = <P^{ik} Z_k, Z_j> = P^{ik} C_{kj}$$

So that $P^{ik}$ is indeed the inverse of $C_{kj}$. Similarly, $<Z^i, Z^j> = P^{ij}$. Since $<Z^i, Z_j> = \delta_j^i$, and since $<Z_i - \mathcal{P}_n^{-i} Z_i, Z_j> = 0$, for all $j \neq i$, then

$$I_i = Z_i - \mathcal{P}_n^{-i} Z_i = \alpha(i) Z^i \text{ for some } \alpha(i) \tag{7}$$

Taking scalar products of both sides with $Z^i$,

$$\Rightarrow 1 = \alpha(i) <Z^i, Z^i>$$

$$\Rightarrow \alpha(i) = \frac{1}{P^{ii}}$$

Using this result and (7)

$$<I_i, I_j> = <Z_i - \mathcal{P}_n^{-i} Z_i, Z_j - \mathcal{P}_n^{-i} Z_j> = \alpha(i)\alpha(j) P^{ik} P^{il} <Z_k, Z_l>$$

$$\text{So } <I_i, I_j> = \frac{P^{ij}}{P^{ii} P^{jj}}$$

Since $\mathcal{P}_n^{-i}$ is an orthogonal projection, it satisfies $\mathcal{P}_n^{-i} Z_i = \min_{Y \in T_n} \|Z_i - Y\|^2$, so it must be the cross-validated kriging value $Z_{-i}^K$ of $Z_i$ and so takes the form $\sum_{j \neq i} \lambda_j Z_j$. But, using (7) again and substituting for $\alpha(i)$

$$I_i = Z_i - Z_{-i}^K = Z_i - \mathcal{P}_n^{-i} Z_i = \frac{Z^i}{P^{ii}} = \frac{\sum_{i=0}^n P^{ij} Z_j}{P^{ii}} = Z_i + \sum_{j \neq i} \frac{P^{ij}}{P^{ii}} Z_j$$

$$\text{So that } Z_{-i}^K = - \sum_{j \neq i} \frac{P^{ij}}{P^{ii}} Z_j$$

and the cross-validation weights are $\lambda_j = -\frac{P^{ij}}{P^{ii}}$. Finally, since the dual vectors are just rescaled innovations (from (7)), and since dual kriging of any $Z(x) \in L^2(\Omega, \mathcal{A}, p)$ takes the form $\sum_i b_i C_{ix}$ where $b_i = P^{ij} Z_j = Z^i$, then

$$Z^K(x) = \sum_i C_{ix} \frac{I_i}{\|I_i\|^2}$$

as required.

*Note 1* Only the case of a positive definite covariance has been considered here. However, the same sort of results can be found for universal kriging and intrinsic random functions. The mechanisms needed for this are well covered in Matheron (1981).

Now turning attention to embedded kriging models in a decision forest, the current analysis will become more heuristic, especially in the case of small data sets. For embedded variables (as well as the data variables), Theorem 2 requires that the embedded variables used for training, which are of the form $\{Z_{-i}^K\}_{i=1}^n$, uniformly sample the full set of kriged values used for estimation of the conditional distributions, $\{Z^K(x); x \in V\}$, i.e. it has the same distribution. While, as discussed earlier, this is true as the number of samples, $n$, tends to infinity for a 'good' sampling scheme, such as randomly spatially sampled, it is not at all true in general for small data sets. As a simple example, consider that our data are on a regular grid separated by a distance just less than or equal to the range of its (stationary) variogram. Then, with 0 mean and simple kriging, all the $Z_{-i}^K = 0$, whereas (almost) all the $Z^K(x)$ are non-zero. Another scenario that can pose problems is when Ember is used for extrapolation. However, let us postpone concern about these cases for now, and look at the case where there are enough samples that the histograms approximately match.

To help sharpen intuition, let us consider the absurd case where the forest is only trained on one strong embedded kriging variable. This would never be recommended in practice, not least because one always has the spatial coordinates available, so that at a minimum they could be used. In general, it appears to be good practice to use at least some additional variables to help randomization between trees, even if they do not appear to be particularly strong variables. However, in this case let us try to estimate $P(Z|Z^K)$. That is, given an unknown Z variable and a kriged estimate for that variable, what is the distribution of Z given $Z^K$. No reference is made to other variables, even spatial location (although they must have been used implicitly to do the kriging). As usual, the forest is trained on the $Z^K_{-i}$. The distribution of Z is found by passing $Z^K$ through each tree in the forest. During tree creation, each node, having only one variable, is such that the split reduces to selecting a split value $z_c$ randomly and moving each training point to the left child if $Z^K_{-i} < z_c$; otherwise move it to the right child. The estimation is passed through the tree using the same stored split values. If, by chance, there happen to be some training values with exactly the same kriged value, they will traverse the tree in exactly the same way, and the prediction will be taken from those samples in the terminal node (so, for estimate of the mean, it will be the average of the samples $Z_i$ whose estimates $Z^K_{-i}$ exactly equal $Z^K$). If the terminal nodes are quite large, there may be some other samples in there, but these will change from one tree to the next, so that the final weights will be dominated by the 'exact samples'. More typically, if $Z^K_{-i} \neq Z^K$, then the probability that they go to different children in a node split is

$$\frac{|Z^K_{-i} - Z^K|}{max_{i,j}|Z^K_{-i} - Z^K_{-j}|}$$

where the denominator is the width of the node. This means that $Z^K$ is more likely to lie in the same node as $Z^K_{-i}$ if $Z^K_{-i} \approx Z^K$, meaning that the forest attributes its high weights preferentially to samples whose cross-validated weight is close to the kriged value of the variable we are trying to estimate (remember, weights are only non-zero for the $i$th sample if it lies in the same terminal node as $Z^K$ for at least one tree). If other variables are used, such as the coordinates, then splits do not invariably depend only on the embedded kriging, and the choice of split depends more strongly on variables which best minimize the in-group variance of the target variable in the children. Weak variables will be chosen less often, and so the terminal node edges will be shorter for strong variables and longer for weak ones, meaning that if kriging is a weak variable, the $Z^K$ may find itself in terminal nodes with values of $Z^K_{-i}$ that are quite different from it. It is noted that the series of decisions about how to split notes are independent variables. Therefore, with non-adaptive splitting strategies it is possible to push the analysis of weights further. For example, splitting at node centres can, with some other simple hypothesis, turn the series of splits into a binomial variable and allow an analysis of the dimension of the edges of $R(y, \theta)$ (e.g. Breiman, 2004). While potentially interesting, this is not further pursued in the current paper, as the hypotheses in that paper are not close to valid for the practical model used here.

To see how closely the mean of the Ember estimate of the distribution resembles kriging, remember that when estimating at a location x, when the secondary variables take the value $Y(x) = y$, then the mean, $\widehat{\mu}_x(y)$, is given by

$$\widehat{\mu}_x(y) = \sum_i \omega_i(y) Z_i = \sum_i \omega_i(Z_i - Z^K_{-i}) + \sum_i \omega_i(Z^K_{-i} - Z^K) + \sum_i \omega_i Z^K,$$

where for convenience the explicit dependence of $\omega_i$ on $y$ has been dropped. Since the sum of weights is 1, then bringing the last term on the right across and taking variance

$$\begin{aligned}
Var\left(\hat{\mu}_x(y) - Z^K\right) &= Var\left(\sum_i \omega_i(Z_i - Z_{-i}^K) + \sum_i \omega_i(Z_{-i}^K - Z^K)\right) \\
&= Var\left(\sum_i \omega_i I_i\right) + \sum_{i,j} \omega_i \omega_j Cov\left(I_i, Z_{-j}^K - Z^K\right) + Var\left(\sum_i \omega_i(Z_{-i}^K - Z^K)\right)
\end{aligned} \quad (9)$$

From (9), under the consistency conditions, in the limit, as n tends to infinity, the Ember estimate of the mean is consistent., i.e., the variance of the LHS tends to 0.

The third term of (9) reflects the discussion about the size of terminal nodes. The weight $\omega_i$ is only non-zero when $Z_{-i}^K$ is in the same terminal node as $Z^K$, $R(y, \theta)$ for some tree, in which case, by Lemma 2, the distance between them is $\leq |J(\theta, y, k)|$, where k is the index of the embedded kriging variable. If $J(y, k) = \max_\theta J(\theta, y, k)$, is the maximum over the finite number of trees, then for all i, $(Z_{-i}^K - Z^K)$ is a bounded random variable taking values in $[-|J(y, k)|, |J(y, k)|] \subset \mathbb{R}$, and it follows from Popoviciu's inequality on variances that $Var(Z_{-i}^K - Z^K) < |J(y, k)|^2$, and in turn, by expanding and applying Cauchy-Schwartz, that $Var(\sum_i \omega_i(Z_{-i}^K - Z^K)) < |J(y, k)|^2$.

Similarly, using Cauchy-Schwartz on the second term of Eq. (9) it follows that $\sum_{i,j} \omega_i \omega_j Cov\left(I_i, Z_{-j}^K - Z^K\right) \leq |J(y, k)| \sum \omega_i \sigma_{-i}$, where $\sigma_{-i} = \sqrt{Var(I_i)}$.

To treat the first term on the right requires an ergodic hypothesis for the innovations ($\rho$-ergodicity is sufficient but seems to be heavy-handed). A simpler approach is to note that in the proof of Theorem 1, based only on the hypotheses 1–3, Meinshausen shows that the sum of squared weights converges to zero. Convergence will typically be faster than if it depended only on the weights since

$$\begin{aligned}
Var\left(\sum_i \omega_i I_i\right) &= \sum_i \omega_i^2 Var(I_i) + \sum_{j \neq i} \omega_i \omega_j Cov\left(I_i, I_j\right) \\
&= \sum_i \omega_i^2 \sigma_{-i}^2 + \sum_{j \neq i} \frac{\omega_i P^{ij} \omega_j}{P^{ii} P^{jj}}.
\end{aligned}$$

Bringing these terms for (9) gives the inequality

$$\begin{aligned}
Var\left(\hat{\mu}_x(y) - Z^K\right) &\leq \sum_i \omega_i^2 \sigma_{-i}^2 + \sum_{j \neq i} \frac{\omega_i P^{ij} \omega_j}{P^{ii} P^{jj}} + |J(y, k)| \sum \omega_i \sigma_{-i} + |J(y, k)|^2 \\
&\approx \sum_i \omega_i^2 \sigma_{-i}^2 + |J(y, k)| \sum \omega_i \sigma_{-i} + |J(y, k)|^2.
\end{aligned}$$

The approximation follows for most practical cases, as in these cases kriging weights will be positive or only slightly negative. But Theorem 3b) shows that cross-validation kriging weights are only negative when $P^{ij} > 0$, so typically (and exactly when the system is positively kriging, i.e. when all kriging weights are positive for all data configurations, which is the case for the class of covariances associated with the resolvent of a stationary reversible Markov chain (Matheron 1986) $Var\left(\sum_i \omega_i I_i\right) \leq \sum_i \omega_i^2 \sigma_{-i}^2$. Hence, convergence of this first term will typically be rapid if kriging has good predictive power, i.e. when $\sigma_{-i}^2 \ll \sigma^2$, but will depend on the convergence of the forest itself (via the sum of squares of weights) when kriging is a poor estimator. This formula gives us a general understanding of when the Ember estimate will be close to the kriging estimate at a location x. Provided there are enough data locations which krige to values $Z_{-j}^K$ close to the kriged value at the target location $Z^K$, then the construction of the forest ensures that the third term is small (and ultimately that the second term is small too). If, moreover, cross-validation works reasonably well at those data locations, then the first term will also be small. In other words, if kriging is a strong variable at x, then the Ember mean will closely resemble the kriged value. If it is not, then Ember will follow the more powerful variables and consequently will take longer to converge towards the kriging estimate (but ultimately will because of Theorem 2).

Two examples of when kriging may be a weak variable in the decision forest were mentioned earlier. The first of those was when the range is shorter than the spacing between data samples. In this case, without loss of generality, assuming the mean is 0, then with simple kriging, $Z_{-i}^K = 0$, whereas $Z^K(x)$ may be non-zero close to data locations. As the experimental variogram resembles the variogram of a pure nugget effect, the forest learns nothing from kriging, which consequently is a weak variable, and the Ember mean does not approximate the kriged value at all locations x. This is as expected, as the envelope should only reflect empirical information. The short range of the hypothetical true variogram kriging has no basis on data alone. It will still be captured in the simulation phase of the Ember model, so its absence in the envelope is irrelevant to the final simulated result. Of course, as the number of data points increases, the number of samples becomes large enough to sample inside the range, and eventually kriging becomes a strong variable.

The second example mentioned was about extrapolation. For example, all the data may be clustered in the centre of a field, but the model is used to extrapolate to a much larger area. In this case, typically the $Z_{-i}^K \neq 0$, and non-zero values will dominate the histogram of the embedded training data, yet in the extrapolated area $Z^K(x) = 0$, so there will be a 'spike' at 0 in the histogram of $Z^K(x)$, meaning condition 1A for Theorem 2 is not satisfied. Convergence is no longer guaranteed in the extrapolated area. That this situation may be problematic is hardly a surprise, as extrapolation is difficult for all geostatistical models, and indeed the notion of convergence is not usually meaningful when extrapolating.

To finish, consider the simple example of having only embedded kriging. The conditional distribution produced by the forest is $\widehat{F}(Z|Y = y) = \sum_{i=1}^n \omega_i(y) \mathbb{I}_{\{Y_i < y\}}$. Returning to the case where kriging is the only explanatory variable in the forest, this

only has non-zero weights when $Z_{-i}^K \approx Z^K$, and the estimate of the envelope distribution is the cumulative histogram of the subset of samples, $Z_i$, whose cross-validation, $Z_{-i}^K$, is close to the kriged value $Z^K$. There is no constraint as to where these sample locations are spatially—the only condition is that they all closely match the specific kriged value $Z^K$. As that value changes, so do the weights, and the predicted cumulative histogram changes as a function of the kriged value (allowing heteroscedasticity to be captured). Suppose, in addition, training uses the spatial coordinates, $x = (x, y, z)$. If at least one coordinate is a strong variable in training, then Ember assigns higher weights to a sample when $Z_{-i}^K \approx Z^K$, *and* where the target location's relevant coordinate is close to the sample's strong coordinate. So, loosely speaking, if $x$ is a strong coordinate but $y$ is not, then samples with similar x coordinate and similar kriged value will be give high weights irrespective of the value of $y$.

For illustration, an example of an Ember model using embedded kriging *only,* or with just coordinates, is shown in Fig. 11. The 'true' image is that used in Example 2. It is worth emphasizing again that this is not at all how the method should be used. It merely helps us understand that embedding kriging can be a strong variable and contribute to producing useful results. The top left is kriging performed after fitting a variogram to the 50 data points, and the simulation with the same variogram is shown below. The two figures in the centre are Ember with embedded kriging and no other variable, and on the right, Ember with the addition of the x and y coordinates in the training. Both kriging and coordinates are weak variables, certainly compared to the earlier versions of this example, which used all the secondary data. The Ember means are a bit washed out, which is to be expected as they are weak variables. The simulations, on the other hand, seem to be slightly better than the pure Gaussian simulation. This would suggest that Ember is already beginning to make some use of the lack of stationarity in the local variability of the true image. This is confirmed in Fig. 12, which has the same configuration for the 800-sample case. Note that both Ember simulations use the same sampling distribution, and are quite similar. Differences are due to the influence of the spatial x,y coordinates.

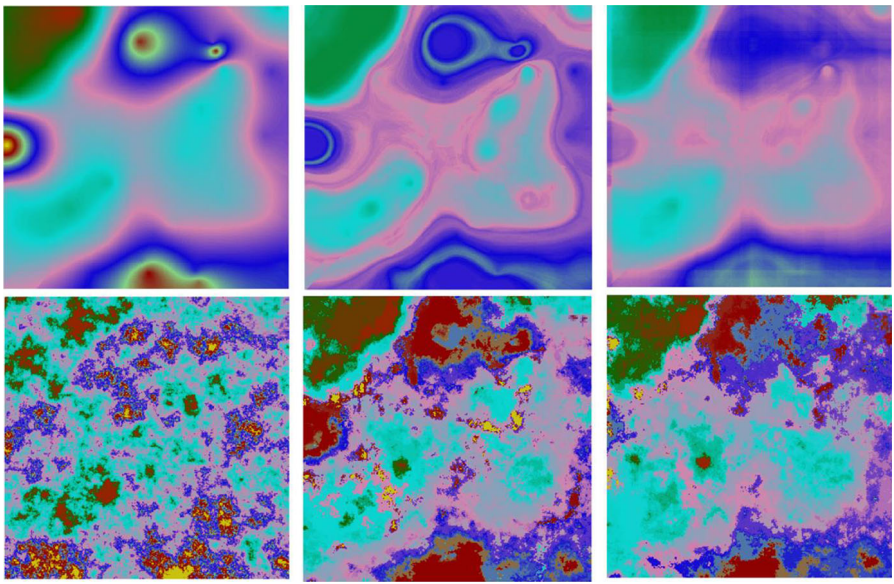

**Fig. 11** Test of usage of embedded model with no other variables, 50-point case. On the left, classic kriging above and Gaussian simulation below. In the centre, estimate and simulation using only embedded kriging (unrealistic, this would never be done in practice). On the right, Ember with embedded kriging and x,y coordinates

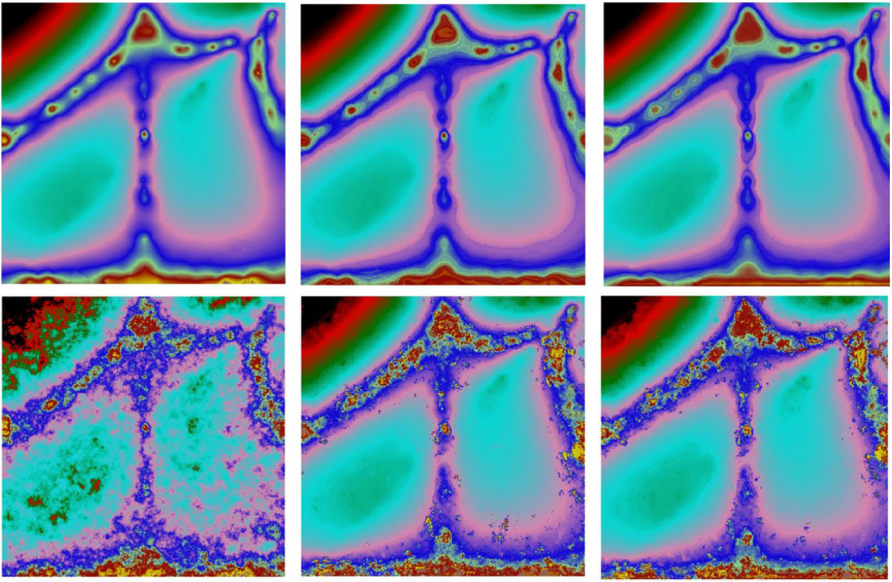

**Fig. 12** Test of usage of embedded model with no other variables. 800-point case. Same configuration of figures as the 50-point case. All three estimates give essentially the same error as measured by MSE, but the two Ember simulations show a gain in terms of quality over the Gaussian simulation by restricting variability only to areas with high uncertainty

The 800-sample case is shown in Fig. 12. It is clear that the embedded model only has already begun to converge to the kriging case. Of course, it brings no advantage over pure kriging, but this is purely to illustrate the argument that convergence does occur.

In terms of quality of results, in the 50-point case, kriging is the best estimator of the three, being approximately 20% closer in MSE to the true result than the other two. But for simulations, the Ember models appear to have the edge in terms of better reproduction of the real texture (though, unsurprisingly with the poor information available, none of the solutions are great). For the 800-point case, the best estimator is Ember with coordinates, followed by kriging and then Ember with only the embedded model, although the difference between them is small. In terms of simulations, it is now clear than the Ember results are significantly better than pure Gaussian simulation. The reason, as has been stated many times now, is that the Gaussian model uses a stationary distribution of the residual 'noise' compared to Ember, which estimates the distribution locally using the envelope. As before, the two Ember simulations use the same sampling random function and so, since kriging is by far the most dominant variable in this case, show only very minor differences between them. Also (not shown), while in the 50-point case, Ember simulations differ from one another substantially all over the image, in the 800-point case the major differences are restricted to the areas near the discontinuities.